# PARTICLE TRANSPORT IN A 3D DUCT BY ADDING AND DOUBLING

B. D. Ganapol[1]
Department of Aerospace and Mechanical
Engineering
University of Arizona

**ABSTRACT**
Particle transport through a duct by Lambertian reflection from duct walls is again considered. This popular transport example has been solved by most numerical transport methods except notably one– the method of doubling. We shall show that the method of doubling provides every bit as, or more, accurate reflectances and transmittances as the numerical discrete ordinates (NDO) and analytical discrete ordinates (ADO) methods with less mathematical and numerical effort.

## INTRODUCTION
In a pivotal paper, Prinja and Pomraning [1] combined their intellectual capital to create the transport equation for neutral particle transport in a duct of any cross sectional area. Assuming classical Lambertian duct wall scattering, neutral particles, such as photons and neutrons move through a duct filled with a nonparticipating medium. Through a clever moments formulation, several investigators, including Ed Larsen, Roberto Garcia and MMR Williams [2-12], investigated solutions to the resulting moments equations via basis functions. In several very informative articles, Garcia [6,11] and co-authors studied the solution for up to three basis functions with conventional numerical discrete ordinates (NDO) and the analytical discrete ordinates (ADO) methods. The difference between the two numerical formulations is in the treatment of the spatial variable with NDO spatially discretizing and ADO forming the analytical spatial solution to the angularly discretized ODEs.

---

[1] Tel/Fax: 520.621.4728/8191
e-mail: ganapol@cowboy.ame.arizona.edu

Here, we again consider the NDO method for transport in a duct, but with a seeming overlooked method of the past found in the radiative transfer literature. Ambartsumian [13] and Chandrasekhar [14] originated the method under the name Principles of Invariance (PoI). PoI is useful for construction of analytical solution representations by solving integral equations. The mathematical setting in star semigroup algebra was presented by Grant and Hunt [15], leading to the derivation of the governing ODEs for slab reflectance and transmittance. A convenient numerical doubling method based on the PoI was then put forth by van de Hulst [16], which found extensive use as a benchmark aiding in the early years of discrete ordinates methods development for solving the radiative transfer equation. In particular, on surveying the early literature, doubling was found to be exceptionally accurate. However, its use seems to have disappeared from the literature with the coming of modern discrete ordinates methods such as DISORT [17] and the various ADO methods of Siewert and co-authors [18-20]. The results of Ref. 21 show doubling with convergence acceleration for the one term duct model to be comparable to the ADO method and simpler in concept. Therefore, with its recent success, it is practical to apply doubling to the 3- basis function model of particle transport in a duct.

The layout of the presentation is as follows. The equations of the GOV model of Ref. 6, named after the authors, will be stated. The response matrix solution to the transport equation then follows. The discretization of the transport equation and its solution, essentially identical to that presented in Ref. 21, is included for completeness. Both isotropic and beam (delta function) sources are considered, where the beam is introduced as an inline delta function through faux quadrature. Next, comes the prescription for the determination of the fundamental slab reflection and transmittance approximations leading to expressions for reflectance and transmittance from the entire slab. With convergence acceleration, we then demonstrate the precision of the doubling formulation and present tables of benchmarks to seven places. We conclude with a discussion of advantages and disadvantages of the doubling method relative to NDO and ADO.

**I. The Discrete Ordinates Balance Equation for the GOV Model**

The *J*-basis function GOV transport equation for the coefficient vector $\boldsymbol{Y}(z,\xi)$ of the basis function representation of the particle distribution [See Ref. 6 for *J* = 3] is

**Table 1.** GOV Pipe Galerkin Models

Underlying model parameters

| | |
|---|---|
| $A = \pi\rho^2$ | $v = \dfrac{8\rho}{3\pi}$ |
| $L = 2\pi\rho$ | $q = 8\rho\left[\dfrac{9\pi}{5}\left(9\pi^2 - 64\right)^{-1/2} - \dfrac{2}{3\pi}\right]$ |
| $u = \dfrac{3\pi}{\rho}\left(9\pi^2 - 64\right)^{-1/2}$ | $r = \rho^{-2}\left[1 - \dfrac{576}{25}\left(9\pi^2 - 64\right)^{-1}\right]^{-1/2}$ |

One basis function ($J = 1$) model

| |
|---|
| $a_{11} = b_{11} = \dfrac{L}{\pi A}$ |

Two basis function ($J = 2$) model

| | |
|---|---|
| $a_{12} = b_{12} = u\left(1 - \dfrac{vL}{\pi A}\right)$ | $a_{21} = b_{21} = -\dfrac{uvL}{\pi A}$ |
| $a_{22} = \dfrac{u^2 v^2 L}{\pi A}$ | $b_{22} = -u^2 v\left(1 - \dfrac{vL}{\pi A}\right)$ |

Three basis function ($J = 3$) model

| | |
|---|---|
| $a_{13} = -qr + \left(v^2 + qv - \dfrac{1}{u^2}\right)\dfrac{rL}{\pi A}$ | $b_{31} = a_{31}$ |
| $a_{23} = \dfrac{2r}{u} - \left(v^2 + qv - \dfrac{1}{u^2}\right)\dfrac{uvrL}{\pi A}$ | $b_{32} = r\left(v^2 + qv - \dfrac{1}{u^2}\right)u\left(1 - \dfrac{vL}{\pi A}\right)$ |
| $a_{33} = \left(v^2 + qv - \dfrac{1}{u^2}\right)\dfrac{r^2 L}{\pi A}$ | $b_{33} = -r\left(v^2 + qv - \dfrac{1}{u^2}\right)\left[qr - \left(v^2 + qv - \dfrac{1}{u^2}\right)\dfrac{rL}{\pi A}\right]$ |
| $a_{31} = \left(v^2 + qv - \dfrac{1}{u^2}\right)\dfrac{rL}{\pi A}$ | $b_{13} = a_{13}$ |
| $a_{32} = -\left(v^2 + qv - \dfrac{1}{u^2}\right)\dfrac{uvrL}{\pi A}$ | $b_{23} = uv\left[qr - \left(v^2 + qv - \dfrac{1}{u^2}\right)\dfrac{rL}{\pi A}\right]$ |

$$\left[\xi\frac{\partial}{\partial z}+A\right]Y(z,\xi)=B\int_{-\infty}^{\infty}d\xi'\psi(\xi')Y(z,\xi') \tag{1a}$$

with kernel

$$\psi(\xi)\equiv\frac{2\omega}{\pi}\frac{1}{\left(1+\xi^2\right)^2} \tag{1b}$$

and boundary conditions

$$Y(0,\xi)=g(\xi)$$
$$Y(Z,-\xi)\equiv 0 \tag{1c}$$

for $\xi \in [0,\infty)$, where the $J$-component vector $Y(z,\xi)$ is

$$Y(z,\xi)\equiv\begin{bmatrix}Y_1 & Y_2 & \ldots & Y_J\end{bmatrix}^T. \tag{1d}$$

$z$ is the spatial variable and $\xi$ (orientation) relates to particle direction. The reflectance and transmittance

$$R_f \equiv \frac{\int_0^\infty d\xi\xi\left(1+\xi^2\right)^{-2}Y_1(0,-\xi)}{\int_0^\infty d\xi\xi\left(1+\xi^2\right)^{-2}g_1(\xi)}$$

$$T_n \equiv \frac{\int_0^\infty d\xi\xi\left(1+\xi^2\right)^{-2}Y_1(Z,\xi)}{\int_0^\infty d\xi\xi\left(1+\xi^2\right)^{-2}g_1(\xi)} \tag{1e,f}$$

are to be determined.

Table 1 gives the three possible incremental basis function models for a duct, which is a pipe of radius $\rho$. Note that the models are nested as the matrices **A** and **B** are filled moving from top to bottom in the table to form matrices of order $J$ for $J$ = 1,2,3. One finds the actual basis functions used and the integrals defining the model parameters in Ref. 6.

### 1. Angular discretization

The basis of the solution to Eqs(1) is the discretization of the scattering integral

$$\int_{-\infty}^{\infty} d\xi' \psi(\xi') Y(z,\xi') = \int_{0}^{\infty} d\xi' \psi(\xi') Y(z,\xi') + \int_{0}^{\infty} d\xi' \psi(\xi') Y(z,-\xi'). \quad (2a)$$

With the (non-unique) change of the $\xi$-variable to

$$\xi = \frac{\mu}{1-\mu} \quad (2b)$$

over the finite range $\mu \in [0,1]$, Eq(2a) becomes

$$\int_{-\infty}^{\infty} d\xi' \psi(\xi') Y(z,\xi') =$$
$$= \int_{0}^{1} d\mu' \frac{d\xi'(\mu')}{d\mu'} \psi(\xi'(\mu')) \Big[ Y(z,\xi'(\mu')) + Y(z,-\xi'(\mu')) \Big], \quad (2c)$$

where

$$\frac{d\xi'}{d\mu'} = \frac{1}{(1-\mu')^2}.$$

In this form, one applies a Gauss/Legendre half range quadrature to give

$$\int_{-\infty}^{\infty} d\xi' \psi(\xi') Y(z,\xi') \simeq \sum_{m'=0}^{N} \omega_{m'} \psi_{m'} \Big[ Y_{m'}^{+}(z) + Y_{m'}^{-}(z) \Big], \quad (3a)$$

where $\mu_m$ are the Gauss-quadrature abscissae from the zeros of the Legendre polynomial of degree $N$

$$P_N(x_m) = 0, \quad x_m \equiv 2\mu_m - 1, \quad m = 1,\ldots,N$$
$$\mu_m = \frac{1}{2}(1 + x_m).$$
(3c,d)

The weights $\omega_m$ are

$$\omega_m = \omega_{N+m} = \frac{1}{(1-\mu_m)^2} v_m$$
(3e)

with [22]

$$v_m = \frac{2(1-x_m^2)}{(N+1)^2 [P_{N+1}(x_m)]^2}.$$
(3f)

Thus, with the discretized $\xi$–variable

$$\xi_m = \frac{\mu_m}{1-\mu_m}, \quad \xi_{N+m} = -\xi_m$$
(4)

introduced into Eq(1a) along with the G/L quadrature, the $N^{th}$ order quadrature approximation for the intensities in the positive (+) and negative (−) orientation of $\xi$

$$\left[\xi_m \frac{\partial}{\partial z} + A\right] Y_m^+(z) \simeq B \sum_{m'=0}^{N} \omega_{m'} \psi_{m'} \left[Y_{m'}^+(z) + Y_{m'}^-(z)\right]$$
$$\left[-\xi_m \frac{\partial}{\partial z} + A\right] Y_m^-(z) \simeq B \sum_{m'=0}^{N} \omega_{m'} \psi_{m'} \left[Y_{m'}^+(z) + Y_{m'}^-(z)\right].$$
(5a,b)

emerges.

With equality, the last expression becomes the $N^{th}$ order discrete ordinates approximation in the continuous spatial variable, where with coefficient vectors $Y_m^\pm(z)$

$$Y_m^\pm(z) \equiv \begin{bmatrix} Y_{1,m}^\pm(z) & Y_{2,m}^\pm(z) & \cdots & Y_{J,m}^\pm(z) \end{bmatrix}^T. \tag{5c}$$

It is convenient to extend the definition of the coefficient vectors by stacking components vertically to give the $JN$ dimensional super-vectors

$$\begin{aligned} Y^+(z) &\equiv \begin{bmatrix} Y_1^T(z) & Y_2^T(z) & \cdots & Y_N^T(z) \end{bmatrix}^T \\ Y^-(z) &\equiv \begin{bmatrix} Y_{N+1}^T(z) & Y_{N+2}^T(z) & \cdots & Y_{2N}^T(z) \end{bmatrix}^T. \end{aligned} \tag{6a}$$

Hence, we obtain the same discrete ordinates vector approximation as in Ref. 21

$$\begin{aligned} \left[\frac{d}{dz} + M^{-1}(D-C)\right] Y^+(z) &= M^{-1} C Y^-(z) \\ \left[-\frac{d}{dz} + M^{-1}(D-C)\right] Y^-(z) &= M^{-1} C Y^+(z); \end{aligned} \tag{6b,c}$$

however, for the super-vectors and the following extended super-matrices of order $J$:

$$\begin{aligned} \psi &\equiv diag\{diag\{\psi_m; j=1,...,J\}; m=1,...,N\} \\ W &\equiv diag\{diag\{\omega_m; j=1,...,J\}; m=1,...,N\} \\ M &\equiv diag\{diag\{\xi_m; j=1,...,J\}; m=1,...,N\}, \\ D &\equiv diag\{\{A\}_m; m=1,...,N\}, \end{aligned} \tag{6d,e,f}$$

and the $N \times N$ partition of the order $J$ matrix $B$

$$B_s \equiv \{\{B\}_{i,j}; i,j=1,...,N\}. \tag{6g}$$

In addition,

$$C \equiv B_s \psi W. \tag{6h}$$

We next consider the numerical treatment of the spatial variable.

## 2. Spatial discretization and single slab response

The numerical approximation to the spatial operator distinguishes the doubling and NDO methods from the ADO method. Doubling proceeds with the discrete approach identical to the NDO method. As soon established, this approach proves to be one of the most straightforward ways of solving 1D transport equations.

If one further defines the full solution vector as

$$\phi(z) \equiv \begin{bmatrix} Y^{T+}(z) & Y^{T-}(z) \end{bmatrix}^T, \tag{7a}$$

Eqs(6b,c) combines to become

$$\frac{d\phi(z)}{dz} + A\phi(z) = 0 \tag{7b}$$

with

$$A = \begin{bmatrix} M^{-1}(D-C) & -M^{-1}C \\ M^{-1}C & -M^{-1}(D-C) \end{bmatrix}. \tag{7c}$$

Since $A$ is constant, the formal solution over a discrete slab $[z_j, z_{j+1}]$ of width $h$, called the fundamental slab, is

$$\phi_{j+1} = e^{-Ah}\phi_j, \tag{8a}$$

where $e^{-Ah}$ is the matrix exponential function. One can express the exponential matrix function in may ways, one of which is analytically through matrix dia-

**Table 2**. Padé Approximants for $e^{-Ah} \simeq P^{*-1}P$

| Padé Approximant | $P^*$ | $P$ |
|---|---|---|
| 0/1 | $[I + hA]^{-1}$ | $I$ |
| 1/1 | $[I + hA/2]^{-1}$ | $[I - hA/2]$ |
| 2/2 | $[I + hA/2 + h^2A^2/12]^{-1}$ | $[I - hA/2 + h^2A^2/12]$ |
| 3/3 | $[I + hA/2 + h^2A^2/10 - h^3A^3/120]^{-1}$ | $[I - hA/2 + h^2A^2/10 + h^3A^3/120]$ |

-gonalization, which corresponds to the ADO method. Here, we choose the discrete approach, where a Padé approximant represents the exponential function,

$$e^{-Ah} \simeq P^{*-1}P. \qquad (8b)$$

Table 2 gives the approximants that we shall consider.

On substitution of a Padé approximant into Eq(8b) and forcing equality

$$P^*\phi_{j+1} = P\phi_j. \qquad (9a)$$

It is this expression, we exploit for the response matrix of the doubling solution.

Figure 1 shows the inputs to $[Y^-_{j+1}, Y^+_j]$, and outputs from $[Y^+_{j+1}, Y^-_j]$ the fundamental slab.

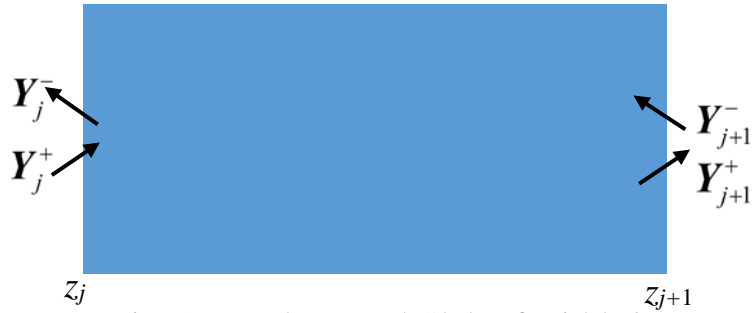

Fig. 1. Fundamental Slab of width $h$.

The task at hand is to find the slab response matrix that when multiplied by the input vector gives the output vector– to be accomplished by partitioning $\boldsymbol{P}^*$ and $\boldsymbol{P}$ into four order *JN* matrices

$$\boldsymbol{P}^* = \begin{bmatrix} \boldsymbol{P}^*_{11} & \boldsymbol{P}^*_{12} \\ \boldsymbol{P}^*_{21} & \boldsymbol{P}^*_{22} \end{bmatrix}$$

$$\boldsymbol{P}_+ = \begin{bmatrix} \boldsymbol{P}_{11} & \boldsymbol{P}_{12} \\ \boldsymbol{P}_{21} & \boldsymbol{P}_{22} \end{bmatrix}$$

(9b,c)

to give for Eq(9a)

$$\begin{bmatrix} \boldsymbol{P}^*_{11} & \boldsymbol{P}^*_{12} \\ \boldsymbol{P}^*_{21} & \boldsymbol{P}^*_{22} \end{bmatrix} \begin{bmatrix} \boldsymbol{Y}^+_{j+1} \\ \boldsymbol{Y}^-_{j+1} \end{bmatrix} = \begin{bmatrix} \boldsymbol{P}_{11} & \boldsymbol{P}_{12} \\ \boldsymbol{P}_{21} & \boldsymbol{P}_{22} \end{bmatrix} \begin{bmatrix} \boldsymbol{Y}^+_j \\ \boldsymbol{Y}^-_j \end{bmatrix}.$$

(9d)

Eq(9d) therefore represents the set of equations

$$\boldsymbol{P}^*_{11}\boldsymbol{Y}^+_{j+1} + \boldsymbol{P}^*_{12}\boldsymbol{Y}^-_{j+1} = \boldsymbol{P}_{11}\boldsymbol{Y}^+_j + \boldsymbol{P}_{12}\boldsymbol{Y}^-_j$$

$$\boldsymbol{P}^*_{21}\boldsymbol{Y}^+_{j+1} + \boldsymbol{P}^*_{22}\boldsymbol{Y}^-_{j+1} = \boldsymbol{P}_{21}\boldsymbol{Y}^+_j + \boldsymbol{P}_{22}\boldsymbol{Y}^-_j,$$

(9e,f)

which upon re-arrangement in terms of input and output vectors become

$$\begin{bmatrix} -\boldsymbol{P}_{12} & \boldsymbol{P}^*_{11} \\ -\boldsymbol{P}_{22} & \boldsymbol{P}^*_{21} \end{bmatrix} \begin{bmatrix} \boldsymbol{Y}^-_j \\ \boldsymbol{Y}^+_{j+1} \end{bmatrix} = \begin{bmatrix} -\boldsymbol{P}^*_{12} & \boldsymbol{P}_{11} \\ -\boldsymbol{P}^*_{22} & \boldsymbol{P}_{21} \end{bmatrix} \begin{bmatrix} \boldsymbol{Y}^-_{j+1} \\ \boldsymbol{Y}^+_j \end{bmatrix}.$$

(9g)

Solving for the output vectors

$$\begin{bmatrix} \boldsymbol{Y}^-_j \\ \boldsymbol{Y}^+_{j+1} \end{bmatrix} = \boldsymbol{R} \begin{bmatrix} \boldsymbol{Y}^-_{j+1} \\ \boldsymbol{Y}^+_j \end{bmatrix}$$

(10a)

yields the desired response matrix (RM) $\boldsymbol{R}$,

$$R \equiv \begin{bmatrix} -P_{12} & P_{11}^* \\ -P_{22} & P_{21}^* \end{bmatrix}^{-1} \begin{bmatrix} -P_{12}^* & P_{11} \\ -P_{22}^* & P_{21} \end{bmatrix}, \tag{10b}$$

provided the inverse exists. As presented, $R$ requires inversion of a $2JN$ order matrix whose effort is to be reduced through the analysis to follow.

### 2.a. Padé Approximant 0/1: Backward Euler

### 2.b. Padé approximant 1/1: Diamond Difference (DD)

Approximant 1/1 is the Diamond Difference (DD) spatial discretization of order $h^2$ of Eqs(7) with

$$P^* \equiv [I + hA/2] = \begin{bmatrix} I + \frac{h}{2}M^{-1}(D-C) & -\frac{h}{2}M^{-1}C \\ \frac{h}{2}M^{-1}C & I - \frac{h}{2}M^{-1}(D-C) \end{bmatrix} \tag{15a}$$

$$P \equiv [I - hA/2] = \begin{bmatrix} I - \frac{h}{2}M^{-1}(D-C) & \frac{h}{2}M^{-1}C \\ -\frac{h}{2}M^{-1}C & I + \frac{h}{2}M^{-1}(D-C) \end{bmatrix}. \tag{15b}$$

If

$$\alpha_+ \equiv I + \frac{h}{2}M^{-1}(D-C)$$

$$\alpha_- \equiv I - \frac{h}{2}M^{-1}(D-C) \tag{15c,d,e}$$

$$\beta \equiv \frac{h}{2}M^{-1}C,$$

then

$$\boldsymbol{P}^* = \begin{bmatrix} \alpha_+ & -\beta \\ \beta & \alpha_- \end{bmatrix} \tag{15f}$$

$$\boldsymbol{P} = \begin{bmatrix} \alpha_- & \beta \\ -\beta & \alpha_+ \end{bmatrix}. \tag{15g}$$

Thus,

$$\begin{bmatrix} -\beta & \alpha_+ \\ -\alpha_+ & \beta \end{bmatrix} \begin{bmatrix} Y_j^- \\ Y_{j+1}^+ \end{bmatrix} = \begin{bmatrix} \beta & \alpha_- \\ -\alpha_- & -\beta \end{bmatrix} \begin{bmatrix} Y_{j+1}^- \\ Y_j^+ \end{bmatrix}. \tag{15h}$$

and on multiplying the bottom equation by $-1$ and flipping the equations

$$\begin{bmatrix} \alpha_+ & -\beta \\ -\beta & \alpha_+ \end{bmatrix} \begin{bmatrix} Y_j^- \\ Y_{j+1}^+ \end{bmatrix} = \begin{bmatrix} \alpha_- & \beta \\ \beta & \alpha_- \end{bmatrix} \begin{bmatrix} Y_{j+1}^- \\ Y_j^+ \end{bmatrix}. \tag{15i}$$

Taking advantage of symmetry by adding and subtracting the two equations together and replacing the original equations

$$\begin{bmatrix} \alpha_+ - \beta & \alpha_+ - \beta \\ \alpha_+ + \beta & -(\alpha_+ + \beta) \end{bmatrix} \begin{bmatrix} Y_j^- \\ Y_{j+1}^+ \end{bmatrix} = \begin{bmatrix} \alpha_- + \beta & \alpha_- + \beta \\ \alpha_- - \beta & -(\alpha_- - \beta) \end{bmatrix} \begin{bmatrix} Y_{j+1}^- \\ Y_j^+ \end{bmatrix},$$

gives the following response matrix:

$$\boldsymbol{R}(1/1) = \begin{bmatrix} \alpha_+ - \beta & \alpha_+ - \beta \\ \alpha_+ + \beta & -(\alpha_+ + \beta) \end{bmatrix}^{-1} \begin{bmatrix} \alpha_- + \beta & \alpha_- + \beta \\ \alpha_- - \beta & -(\alpha_- - \beta) \end{bmatrix}. \tag{15j}$$

On multiplication of the partitioned matrices, and continuing by explicitly representing the inverse in Eq(15j), there results

$$\boldsymbol{R}(1/1) = \begin{bmatrix} \boldsymbol{T}_n & \boldsymbol{R}_f \\ \boldsymbol{R}_f & \boldsymbol{T}_n \end{bmatrix}, \tag{15k}$$

where

$$\begin{Bmatrix} T_n \\ R_f \end{Bmatrix} = \frac{1}{2}\left[(\alpha_+ - \beta)^{-1}(\alpha_- + \beta) \pm (\alpha_+ + \beta)^{-1}(\alpha_- - \beta)\right]. \qquad (15l)$$

Note that Eq (15k) is the universal symmetric representation of the response matrix since it states that reflection and transmission are independent of which surface particles enter.[B1]

### 2.c. Padé approximant 2/2: Diamond Difference Modification 1 (DDM1)

A similar pattern holds for the last two approximants, which are variatiants of DD. For DDM1 of order $h^3$, one confirms the pattern since $A$ of Eq(7c) on partitioning is

$$A = \begin{bmatrix} A_{11} & A_{12} \\ -A_{12} & -A_{11} \end{bmatrix}$$

and squared gives the symmetric matrix

$$A^2 = \begin{bmatrix} A_{11}^2 - A_{12}^2 & A_{11}A_{12} - A_{21}A_{11} \\ -A_{21}A_{11} + A_{11}A_{12} & A_{11}^2 - A_{12}^2 \end{bmatrix} \equiv \begin{bmatrix} \gamma_{11} & \gamma_{12} \\ \gamma_{12} & \gamma_{11} \end{bmatrix}. \qquad (16a)$$

Therefore, from Eqs(9b,c) and the 2/2 approximant

$$P^* = \begin{bmatrix} \alpha_+ + \dfrac{h^2}{12}\gamma_{11} & -\left(\beta - \dfrac{h^2}{12}\gamma_{12}\right) \\ \beta + \dfrac{h^2}{12}\gamma_{12} & \alpha_- + \dfrac{h^2}{12}\gamma_{11} \end{bmatrix} \qquad (16b)$$

$$P = \begin{bmatrix} \alpha_- + \dfrac{h^2}{12}\gamma_{11} & \beta + \dfrac{h^2}{12}\gamma_{12} \\ -\left(\beta - \dfrac{h^2}{12}\gamma_{12}\right) & \alpha_+ + \dfrac{h^2}{12}\gamma_{11} \end{bmatrix}. \qquad (16c)$$

To see the pattern, let

$$\alpha_+ \to \alpha_+ + \dfrac{h^2}{12}\gamma_{11}$$

$$\alpha_- \to \alpha_- + \dfrac{h^2}{12}\gamma_{11}$$

$$\beta_+ \equiv \beta + \dfrac{h^2}{12}\gamma_{12} \qquad (16d,e,f,g)$$

$$\beta_- \equiv \beta - \dfrac{h^2}{12}\gamma_{12}$$

and following the procedure for approximant 1/1, we find

$$\boldsymbol{R}(2/2) = \begin{bmatrix} \alpha_+ - \beta_+ & \alpha_+ - \beta_+ \\ \alpha_+ + \beta_+ & -(\alpha_+ + \beta_+) \end{bmatrix}^{-1} \begin{bmatrix} \alpha_- + \beta_- & \alpha_- + \beta_- \\ \alpha_- - \beta_- & -(\alpha_- - \beta_-) \end{bmatrix} \qquad (16h)$$

and

$$\begin{Bmatrix} T_n \\ R_f \end{Bmatrix} = \dfrac{1}{2}\left[(\alpha_+ - \beta_+)^{-1}(\alpha_- + \beta_-) \pm (\alpha_+ + \beta_+)^{-1}(\alpha_- - \beta_-)\right]. \qquad (16i)$$

Eq(15k) therefore becomes

$$\boldsymbol{R}(2/2) = \begin{bmatrix} T_n & R_f \\ R_f & T_n \end{bmatrix}. \qquad (16j)$$

**2.d. Padé approximant 2/2: Diamond Difference Modification 2 (DDM2)**
The pattern also holds similarly for DDM2 since

$$A^3 = \begin{bmatrix} \gamma_{11}A_{11} - \gamma_{12}A_{12} & \gamma_{11}A_{12} - \gamma_{12}A_{11} \\ -(\gamma_{11}A_{12} - \gamma_{12}A_{11}) & -(\gamma_{11}A_{11} - \gamma_{12}A_{12}) \end{bmatrix} \equiv \begin{bmatrix} m_{11} & m_{12} \\ -m_{12} & -m_{11} \end{bmatrix}. \quad (17a)$$

Then defining

$$\alpha_+^{+-} \equiv \alpha_+ + \frac{h^2}{10}\gamma_{11} - \frac{h^3}{120}m_{11}$$

$$\alpha_-^{++} \equiv \alpha_- + \frac{h^2}{10}\gamma_{11} + \frac{h^3}{120}m_{11}$$

$$\beta^{-+} \equiv \beta + \frac{h^2}{10}\gamma_{12} - \frac{h^3}{120}m_{12}$$

$$\beta^{++} \equiv \beta + \frac{h^2}{10}\gamma_{12} + \frac{h^3}{120}m_{12},$$

(17bc,d,e)

one can show, as above, that

$$\begin{Bmatrix} T_n \\ R_f \end{Bmatrix} = \frac{1}{2}\left[ \left(\alpha_+^{+-} - \beta^{++}\right)^{-1}\left(\alpha_-^{++} + \beta^{-+}\right) \pm \left(\alpha_+^{+-} + \beta^{++}\right)^{-1}\left(\alpha_-^{++} - \beta^{-+}\right) \right]. \quad (17f)$$

and

$$R(3/3) = \begin{bmatrix} T_n & R_f \\ R_f & T_n \end{bmatrix}. \quad (17g)$$

We now focus our attention to the entire slab to build the slab response matrix.

**Analytical Response matrix**

$$\begin{bmatrix} x^- & -B \\ -B & x^- \end{bmatrix} \begin{bmatrix} Y_j^- \\ Y_{j+1}^+ \end{bmatrix} = \begin{bmatrix} B & x^+ \\ x^+ & B \end{bmatrix} \begin{bmatrix} Y_{j+1}^- \\ Y_j^+ \end{bmatrix}$$

$$x^{\pm} \equiv \alpha + \beta \pm A$$

$$A \equiv \left.\frac{d\boldsymbol{H}(h-z)}{dz}\right|_{z=0} = \boldsymbol{T}\left[diag\left\{-\frac{\lambda_k \cosh(\lambda_k h)}{\sinh(\lambda_k h)}\right\}\right]\boldsymbol{T}^{-1} = -\sqrt{\tilde{A}}\ coth\left(\sqrt{\tilde{A}}h\right)$$

$$B \equiv \left.\frac{d\boldsymbol{H}(z)}{dz}\right|_{z=0} = \boldsymbol{T}\left[diag\left\{\frac{\lambda_k}{\sinh(\lambda_k h)}\right\}\right]\boldsymbol{T}^{-1} = \sqrt{\tilde{A}}\ csch\left(\sqrt{\tilde{A}}h\right)$$

$$\tilde{A} \equiv (\alpha + \beta)(\alpha - \beta)$$

$$\alpha \equiv M^{-1}(D-C)$$

$$\beta \equiv M^{-1}C$$

$$R \equiv \begin{bmatrix} x^- & -B \\ -B & x^- \end{bmatrix}^{-1} \begin{bmatrix} B & x^+ \\ x^+ & B \end{bmatrix}$$

$$\begin{Bmatrix} R_f \\ T_n \end{Bmatrix} = \frac{1}{2}\left[(x^- - B)^{-1}(x^+ + B) \pm (x^- + B)^{-1}(x^+ - B)\right]$$

$$U\ coth(U) = I + \sum_{n=1}^{\infty} \frac{2^{2n}}{(2n!)} B_{2n}\ U^{2n}$$

$$U\ csch(U) = I + \sum_{n=1}^{\infty} \frac{(2-2^{2n})}{(2n!)} B_{2n}\ U^{2n}$$

$$x^{\pm} \equiv \alpha + \beta \pm A$$

$$\begin{Bmatrix} R_f \\ T_n \end{Bmatrix} = \frac{1}{2}\left[\begin{matrix}(h(\alpha+\beta) + h(-A-B))^{-1}(h(\alpha+\beta) - h(-A-B)) \pm \\ \pm (h(\alpha+\beta) + h(-A+B))^{-1}(h(\alpha+\beta) - h(-A+B))\end{matrix}\right]$$

$$h(-\mathbf{A}-\mathbf{B}) = 2\sum_{n=1}^{\infty}\frac{(2^{2n}-1)}{(2n!)}B_{2n}\,\tilde{\mathbf{A}}^n h^{2n}$$

$$h(-\mathbf{A}+\mathbf{B}) = 2\mathbf{I} + 2\sum_{n=1}^{\infty}\frac{2^{2n}}{(2n!)}B_{2n}\,\tilde{\mathbf{A}}^n h^{2n}$$

## II. Adding and Doubling Slab Responses

In summary, for a discrete input angular intensity on the fundamental slab $j$ of width $h$, the discrete ordinates balance equation for the output angular intensity is

$$\begin{bmatrix}\mathbf{Y}_j^-\\ \mathbf{Y}_{j+1}^+\end{bmatrix} = \mathbf{R}\begin{bmatrix}\mathbf{Y}_{j+1}^-\\ \mathbf{Y}_j^+\end{bmatrix}, \qquad (18)$$

with input and output linked through the response matrix $\mathbf{R}$. The above determination of $\mathbf{R}$ is not unique, since infinitesimal generators like those of Evans and Stephens [24] or matrix exponentials of Watterman [25] and Liu and Weng [26] or the continuous form of Twomey [27] or the original first collision expressions of van der Hulst and Hansen [16, 28] are also possible. The four response matrix forms proposed above are believed to be more straightforward than those cited however.

In the section, we take a step closer to determining the cumulative response for the full homogeneous slab by finding the response of two slabs added together.

### 1. Adding a single slab

The doubling process begins by adding two slab responses. Consider two adjacent slabs as shown in Fig. 2. For purposes of explanation, the first slab can be a heterogeneous composite of $l-1$ slabs of any widths; while, the second is the fundamental homogeneous slab of width $h$ whose response is $R$ of Eq(10b). Let the response of the heterogeneous slab be $Q_{l-1}$ yielding the exiting intensities

$$\begin{bmatrix} Y_0^- \\ Y_{l-1}^+ \end{bmatrix} = Q_{l-1} \begin{bmatrix} Y_{l-1}^- \\ Y_0^+ \end{bmatrix}. \tag{19a}$$

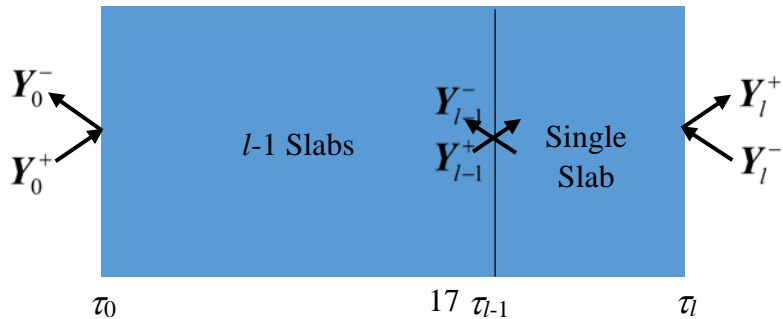

Fig. 2. Two adjacent slabs– the first is heterogeneous composed of $l-1$ slabs and the second is a single homogeneous slab.

In terms of the four order $JN$ partitions, $Q_{l-1}$ is

$$Q_{l-1} = \begin{bmatrix} Q_{l-1,1} & Q_{l-1,2} \\ Q_{l-1,2} & Q_{l-1,1} \end{bmatrix}. \tag{19b}$$

Recall that for the single slab as above

$$\begin{bmatrix} Y_{l-1}^- \\ Y_l^+ \end{bmatrix} = R \begin{bmatrix} Y_l^- \\ Y_{l-1}^+ \end{bmatrix}. \tag{19c}$$

In expanded form, these partitioned equations become

$$\begin{aligned} Y_0^- &= Q_{l-1,1} Y_{l-1}^- + Q_{l-1,2} Y_0^+ \\ Y_l^+ &= R_f Y_l^- + T_n Y_{l-1}^+ \\ Y_{l-1}^+ &= Q_{l-1,2} Y_{l-1}^- + Q_{l-1,1} Y_0^+ \\ Y_{l-1}^- &= T_n Y_l^- + R_f Y_{l-1}^+. \end{aligned} \tag{20a,b,c,d}$$

From the last two equations,

$$\begin{bmatrix} I & -Q_{l-1,2} \\ -R_f & I \end{bmatrix} \begin{bmatrix} Y_{l-1}^+ \\ Y_{l-1}^- \end{bmatrix} = \begin{bmatrix} 0 & -Q_{l-1,1} \\ -T_n & 0 \end{bmatrix} \begin{bmatrix} Y_l^- \\ Y_0^+ \end{bmatrix},$$ (20e)

and from re-arrangement of the first two equations,

$$\begin{bmatrix} Y_0^- \\ Y_l^+ \end{bmatrix} = \begin{bmatrix} 0 & Q_{l-1,2} \\ R_f & 0 \end{bmatrix} \begin{bmatrix} Y_l^- \\ Y_0^+ \end{bmatrix} + \begin{bmatrix} 0 & Q_{l-1,1} \\ T_n & 0 \end{bmatrix} \begin{bmatrix} Y_{l-1}^+ \\ Y_{l-1}^- \end{bmatrix},$$ (20f)

and on inversion of Eq(20e),

$$\begin{bmatrix} Y_{l-1}^+ \\ Y_{l-1}^- \end{bmatrix} = U_{l-1} \begin{bmatrix} Y_l^- \\ Y_0^+ \end{bmatrix}$$ (21a)

with

$$U_{l-1} \equiv w_{m,l-1}^{-1} w_{p,l-1}$$

$$w_{m,l-1} \equiv \begin{bmatrix} I & -Q_{l-1,2} \\ -R_f & I \end{bmatrix}$$ (21b,c,d)

$$w_{p,l-1} \equiv \begin{bmatrix} 0 & Q_{l-1,1} \\ T_n & 0 \end{bmatrix}.$$

When Eq(21a) is introduced into Eq(20f), there results

$$\begin{bmatrix} Y_0^- \\ Y_l^+ \end{bmatrix} = \left\{ \begin{bmatrix} 0 & Q_{l-1,1} \\ T_n & 0 \end{bmatrix} U_{l-1} + \begin{bmatrix} 0 & Q_{l-1,2} \\ R_f & 0 \end{bmatrix} \right\} \begin{bmatrix} Y_l^- \\ Y_0^+ \end{bmatrix},$$ (22a)

from which the combined response (a recurrence in the partitioned matrices),

$$Q_l = \begin{bmatrix} 0 & Q_{l-1,1} \\ T_n & 0 \end{bmatrix} U_{l-1} + \begin{bmatrix} 0 & Q_{l-1,2} \\ R_f & 0 \end{bmatrix},$$ (22b)

follows. To complete the recurrence, an explicit representation for $U_l$ comes from Eq(21b) with $l$ incremented

$$\begin{bmatrix} I & -Q_{l,2} \\ -R_f & I \end{bmatrix} U_l = \begin{bmatrix} 0 & Q_{l,1} \\ T_n & 0 \end{bmatrix} \tag{23a}$$

by multiplying the top equation by $R_f$, adding the result to the bottom equation and replacing the bottom equation with the new equation to give

$$\begin{bmatrix} I & -Q_{l,2} \\ 0 & I - R_f Q_{l,2} \end{bmatrix} U_l = \begin{bmatrix} 0 & Q_{l,1} \\ T_n & R_f Q_{l,1} \end{bmatrix}. \tag{23b}$$

From Schur's complement [23] therefore,

$$U_l = \begin{bmatrix} U_{l,1} & U_{l,2} \\ U_{l,3} & U_{l,4} \end{bmatrix}$$

$$= \begin{bmatrix} Q_{l,2}\left[I - R_f Q_{l,2}\right]^{-1} T_n & Q_{l,1} + Q_{l,2}\left[I - R_f Q_{l,2}\right]^{-1} R_f Q_{l,1} \\ \left[I - R_f Q_{l,2}\right]^{-1} T_n & \left[I - R_f Q_{l,2}\right]^{-1} R_f Q_{l,1} \end{bmatrix} \tag{23c}$$

and therefore decrementing

$$U_{l-1} =$$
$$= \begin{bmatrix} Q_{l-1,2}\left[I - R_f Q_{l-1,2}\right]^{-1} T_n & Q_{l-1,1} + Q_{l,2}\left[I - R_f Q_{l-1,2}\right]^{-1} R_f Q_{l-1,1} \\ \left[I - R_f Q_{l-1,2}\right]^{-1} T_n & \left[I - R_f Q_{l-1,2}\right]^{-1} R_f Q_{l-1,1} \end{bmatrix}. \tag{23d}$$

Thus,

$$\begin{bmatrix} Y_0^- \\ Y_l^+ \end{bmatrix} = Q_l \begin{bmatrix} Y_l^- \\ Y_0^+ \end{bmatrix}, \tag{24a}$$

with the following recurrence for $Q_l$:

$$Q_l = \begin{bmatrix} Q_{l,1} & Q_{l,2} \\ Q_{l,3} & Q_{l,4} \end{bmatrix} = \begin{bmatrix} Q_{l-1,1}U_{l-1,3} & Q_{l-1,1}U_{l-1,4} + Q_{l-1,2} \\ T_n U_{l-1,1} + R_f & T_n U_{l-1,2} \end{bmatrix} \quad (24b)$$

for two slabs together and $U_{l-1}$ from Eq(23c).

## 2. Doubling slab responses

Since one expects the interval $h$ to be small to obtain extreme accuracy for the reflection and especially for the transmission, it seems foolish to continually add slabs one at a time as suggested by Eqs(24). However, this is exactly what one does iteratively in the NDO method. Instead, we apply doubling.

Let the homogeneous medium of width $Z$ be partitioned into $2^n$ subintervals $h_n$, where $n$ is a positive integer and $h_n \equiv Z/2^n$ is the fundamental interval. If the first interval is denoted 1, let $R$, from Eq(10b), be the response for that interval. Next, find the combined response for two such slabs from Eqs(24b), (23d) by setting $Q_{l-1}$ to $R$. Now a slab of width $2h_n$ has the known response, called $Q_2$. Then, double this response by combining the responses of two such slabs of width $2h_2$ again using Eqs(24b), (23d) with $Q_{l-1}$ and $R$ replaced by $Q_2$. Now a slab of width $4h_2$ has the response $Q_3$. This continues until the entire slab is covered to give $Q_n$.

## 2.a. The exiting intensities

Once $Q_n$ is known, the exiting intensity vector from the composite slab of $2^n$ subintervals is

$$\begin{bmatrix} Y_0^- \\ Y_n^+ \end{bmatrix} = Q_n \begin{bmatrix} Y_n^- \\ Y_0^+ \end{bmatrix}, \quad (25)$$

since one knows the incoming angular intensity vector $\begin{bmatrix} Y_n^- & Y_0^+ \end{bmatrix}^T$. We then find the slab reflectance and transmittance in the discrete ordinates approximation from Eq(1e,f) as

$$R_f(n,N) = \left\{ \begin{bmatrix} e_1^T & e_2^T & \cdots & e_{3N}^T \end{bmatrix}^T MW\psi Y^-(0) \right\} / \left\{ \begin{bmatrix} e_1^T & e_2^T & \cdots & e_{3N}^T \end{bmatrix}^T MW\psi g \right\}$$

$$T_n(n,N) = \left\{ \begin{bmatrix} e_1^T & e_2^T & \cdots & e_{3N}^T \end{bmatrix}^T MW\psi Y^+(Z) \right\} / \left\{ \begin{bmatrix} e_1^T & e_2^T & \cdots & e_{3N}^T \end{bmatrix}^T MW\psi g \right\}$$

(26a,b)

where $e_m^T$ is the $J-$ component vector

$$e_m^T \equiv \begin{bmatrix} 1 & 0 & \cdots & 0 \end{bmatrix}^T.$$

Note that the dependence on $n$ and $N$ have be indicated explicitly.

### 2.b. Determination of response for combination of $2^{l-1}$ slabs with $2^{l-1}$ slabs

We end this section with the determination of the response $Q_l$ for the combination of $2^{l-1}$ slabs with $2^{l-1}$ slabs since this response is basic for doubling. We assume all slabs identical.

Since the response for $2^{l-1}$ slabs is $Q_{l-1}$, let $R_f = Q_{l-1,2}$ and $T_n = Q_{l-1,1}$ in Eq(22b) to give

$$Q_l = \begin{bmatrix} 0 & Q_{l-1,1} \\ Q_{l-1,1} & 0 \end{bmatrix} U_{l-1} + \begin{bmatrix} 0 & Q_{l-1,2} \\ Q_{l-1,2} & 0 \end{bmatrix};$$

(27a)

and similarly in Eq(21c,d)

$$w_{m,l-1} \equiv \begin{bmatrix} I & -Q_{l-1,2} \\ -Q_{l-1,2} & I \end{bmatrix}$$

$$w_{p,l-1} \equiv \begin{bmatrix} 0 & Q_{l-1,1} \\ Q_{l-1,1} & 0 \end{bmatrix}$$

(27b,c)

Therefore, from Eq(21b)

$$\begin{bmatrix} I & -Q_{l-1,2} \\ -Q_{l-1,2} & I \end{bmatrix} U_{l-1} \equiv \begin{bmatrix} 0 & Q_{l-1,1} \\ Q_{l-1,1} & 0 \end{bmatrix}. \qquad (27d)$$

By adding and subtracting the bottom equation to and from the top and replacing the original set by the resulting two equations, there results

$$\begin{bmatrix} I-Q_{l-1,2} & I-Q_{l-1,2} \\ I+Q_{l-1,2} & -(I+Q_{l-1,2}) \end{bmatrix} U_l \equiv \begin{bmatrix} Q_{l-1,1} & Q_{l-1,1} \\ -Q_{l-1,1} & Q_{l-1,1} \end{bmatrix}. \qquad (27e)$$

Inverting gives

$$\begin{Bmatrix} U_{l-1,1} = U_{l-1,4} \\ U_{l-1,2} = U_{l=1,3} \end{Bmatrix} = \frac{1}{2} \left\{ [I-Q_{l-1,2}]^{-1} \mp [I+Q_{l-1,2}]^{-1} \right\} Q_{l-1,1} \qquad (27f)$$

and consequently from Eq(27a)

$$Q_l = \begin{bmatrix} Q_{l-1,1} U_{l-1,2} & Q_{l-1,1} U_{l-1,1} + Q_{l-1,2} \\ Q_{l-1,1} U_{l-1,1} + Q_{l-1,2} & Q_{l-1,1} U_{l-1,2} \end{bmatrix}, \qquad (27g)$$

where the recurrence by partition begins with $Q_{1,2} = R_f$, and $Q_l$ now represents the response for the composite of $2^{l-1}$ identical slabs.

**III. Demonstration of solution by doubling**

In this section, we discuss the implementation of the Response Matrix Converged Accelerated Doubling (RMCAD) method. Central to the method will be convergence acceleration. A part of the implementation process is to determine which of the four proposed response matrix approximations is best based on precision and CPU time. All computations are performed on an HP ENVY m4 2.4GHz Notebook. Note that we expect all digits presented to be correct to one unit in the last place unless stated otherwise.

**1. Convergence acceleration**

Our primary interest is in the determination of the reflectance $R_f$ and transmittance $T_n$ from Eqs(26a,b) though exiting angular intensity is a byproduct and is recoverable if desired. To achieve our goal of the highest possible precision for the least effort, we apply convergence acceleration in the forms of Richardsons and the Wynn-epsilon (*W-e*) extrapolations [29].

Implementation of the RMCAD method goes as follows. After determination of the abscissae $\mu_m$, $m = 1,…,N$ for a quadrature order *N*, and partitioning of the slab into $2^n$ subintervals $h_n$, $Q_n$ is found by the doubling procedure described above. From $Q_n$, we find the outgoing angular intensity vector at the abscissae on the slab surfaces from Eq(25) from which $R_f$ and $T_n$ are found from Eqs(26a,b).

What was just described is one solution in a sequence of solutions, where the coordinate (*n,N*) defines each element by number of doublings and quadrature order. The sequence elements follow from incrementing *n* and *N* by strides of one and two starting at 2

$n = 2,…,n_l$; $N=2,4,…,N_L$.

The implementation first converges spatial doubling for each *N* and then increments to the next quadrature order until the sequences for both $R_f$ and $T_n$ converges. To aid convergence, convergence acceleration applies to the inner doubling sequence, embedded in the outer quadrature sequence to which acceleration also applies. Note that the acceleration does not disturb the original sequence and generates a second "accelerated" sequence. The original and accelerated sequences compete for convergence with the first to converge declared the winner. Both accelerations apply to the spatial sequences and only *W-e* applies to quadrature convergence since the sequence does not have a regular error representation, as required for Richardsons extrapolation. Also note that one treats convergence of the reflectance and transmittance sequences independently.

Convergence acceleration, for readers not familiar with the concept, is the replacement of an original sequence by a more rapidly convergent one, which converges to the same limit. There are many ways of developing such a sequence [29] and here we will use a linear, Richardson extrapolation, and non- linear, Wynn-epsilon extrapolation. While extrapolations are not always guaranteed to accelerate convergence, at worst they may give the original rate of convergence but seldom converge to an incorrect limit.

In the following examples and demonstrations, we consider only a circular pipe duct for the $J = 3$ transport model. For a pipe, the spatial variable scales as $Z/\rho$.

As an example of spatial convergence, we consider a pipe of length $Z = 1$, for no absorption (the conservative case, $\omega = 1$) and uniform isotropic incidence

$$Y(0,\xi) = \begin{bmatrix} 1 & 0 & 0 \end{bmatrix}^T.$$

The backward Euler response of Eq(14d) is chosen to introduce the figures of merit of the RMCAD method. Tables 3a and 3b give the spatial doubling and angular quadrature convergence profiles up to– and at– convergence respectively. Note that here convergence means simple "engineering convergence" or Cauchy convergence, where the relative error $\varepsilon_r$ between the last two sequence elements is within a user specified limit, which we set to $10^{-9}$. The figures of merit are the precision, number of doublings $n_c$ and the quadrature order $N_c$ when $R_f$ and $T_n$ have converged to the desired relative error. The mode of theconvergence sequences, original or accelerated, is also of interest as is the CPU time of computation.

Table 3a shows the progression of convergence through doubling, $n$, at the converged quadrature order $N = 34$ (in the second column) for the BE response matrix approximation of $R_f$ and $T_n$. Columns 3 to 8 contain the ratio of the relative error modes of the original (*ORI*), Wynn-epsilon (*W-e*) and Richardsons (*R*) accelerated reflectance and transmittance to the minimum relative error. Other than the first two rows, which are identical for all three modes since two elements of the sequence are required to initiate the accelerations, an entry of unity indicates the converged mode of the sequence. Thus, for this example, Richardsons acceleration converges first. The advantage of convergence acceleration is shown by the corresponding entries for the other modes in the same row, indicating the factor by which that mode relative error exceeds the converged one. Note that in the previous row, *W-e* was the most accurate, but the relative error was not below $10^{-9}$. In some cases, the original is the most accurate. Thus, for this example, spatial convergence requires $n_c = 18$ doublings and quadrature order $N_c = 34$. The slab width is therefore $h_{18} = 1/2^{18} = 3.81 \times 10^{-6}$. The last column is the sum $R_f + T_n$ and should be unity and is for more than 6–places except for the initial doublings, which are hopelessly inaccurate for the BE approximation.

Table 3b shows the convergence in quadrature order $N$. The first column is the number of doublings for the run-up to convergence at $N = 34$ (given in the second column) progressing towards convergence. One finds the ratio of the relative errors of the original reflectance and transmittance to that of the $W$-$e$ acceleration in the following columns. If the entry is larger than unity, then the precision of the $W$-$e$ sequence is greater than the original. Towards convergence, $W$-$e$ has the advantage. Thus, convergence acceleration proceeds faster to convergence than the original both for doubling and quadrature approximation for the BE response matrix approximation.

**Table 3a**. BE response matrix: Doubling to convergence at $N = 34$

| | | $R_f$ | | | $T_n$ | | | |
|---|---|---|---|---|---|---|---|---|
| $n$ | $N$ | $\varepsilon(ORI)/\varepsilon_{min}$ | $\varepsilon/(W\text{-}e)/\varepsilon_{min}$ | $\varepsilon(R)/\varepsilon_{min}$ | $\varepsilon(ORI)/\varepsilon_{min}$ | $\varepsilon(W\text{-}e)/\varepsilon_{min}$ | $\varepsilon(R)/\varepsilon_{min}$ | $R_f+T_n$ |
| 2 | 34 | 1.00E+00 | 1.00E+00 | 1.00E+00 | 1.00E+00 | 1.00E+00 | 1.00E+00 | 1.000000E+00 |
| 3 | 34 | 1.00E+00 | 1.00E+00 | 1.00E+00 | 1.00E+00 | 1.00E+00 | 1.00E+00 | 1.000031E+00 |
| 4 | 34 | 1.00E+00 | 1.10E+02 | 1.01E+00 | 1.00E+00 | 1.10E+02 | 1.01E+00 | -6.733196E+02 |
| 5 | 34 | 1.00E+00 | 1.05E+00 | 1.00E+00 | 1.00E+00 | 1.00E+00 | 1.00E+00 | -1.072139E+09 |
| 6 | 34 | 2.18E+04 | 1.37E+01 | 1.00E+00 | 7.55E+04 | 4.15E+01 | 1.00E+00 | -8.192482E+09 |
| 7 | 34 | 7.18E+02 | 1.00E+00 | 3.98E+00 | 6.97E+02 | 1.00E+00 | 3.99E+00 | 8.959852E+08 |
| 8 | 34 | 3.26E+03 | 1.00E+00 | 2.43E+01 | 1.09E+02 | 1.00E+00 | 2.41E+01 | 6.720107E+08 |
| 9 | 34 | 1.00E+00 | 9.86E+08 | 1.78E+01 | 1.00E+00 | 3.37E+09 | 3.89E+01 | 1.000000E+00 |
| 10 | 34 | 1.00E+00 | 1.00E+00 | 8.42E+02 | 1.00E+00 | 1.00E+00 | 1.73E+03 | 1.000000E+00 |
| 11 | 34 | 1.00E+00 | 1.04E+00 | 4.46E+04 | 1.00E+00 | 1.04E+00 | 9.16E+04 | 1.000000E+00 |
| 12 | 34 | 1.00E+00 | 2.20E+01 | 4.74E+07 | 1.00E+00 | 2.20E+01 | 9.85E+07 | 1.000000E+00 |
| 13 | 34 | 1.00E+00 | 1.02E+00 | 2.02E+08 | 1.00E+00 | 1.02E+00 | 1.93E+08 | 1.000000E+00 |
| 14 | 34 | 1.03E+00 | 1.00E+00 | 8.46E+05 | 1.03E+00 | 1.00E+00 | 2.98E+06 | 1.000000E+00 |
| 15 | 34 | 1.00E+00 | 1.87E+00 | 3.25E+03 | 1.00E+00 | 1.87E+00 | 1.15E+04 | 1.000000E+00 |
| 16 | 34 | 5.84E+00 | 1.00E+00 | 2.46E+01 | 5.92E+00 | 1.00E+00 | 1.22E+02 | 1.000000E+00 |
| 17 | 34 | 4.09E+01 | 1.00E+00 | 3.43E+00 | 4.06E+01 | 1.00E+00 | 2.74E+00 | 1.000000E+00 |
| 18 | 34 | 1.97E+02 | 5.62E+00 | 1.00E+00 | 1.93E+03 | 2.72E+01 | 1.00E+00 | 1.000000E+00 |

**Table 3b**. BE response matrix: Run-up to quadrature order convergence

| $n$ | $N$ | $\varepsilon(ORI)/\varepsilon(W\text{-}e)$ | $\varepsilon(ORI)/\varepsilon(W\text{-}e)$ |
|---|---|---|---|
| 12 | 2 | 1.00E+00 | 1.00E+00 |
| 13 | 4 | 1.00E+00 | 1.00E+00 |
| 14 | 6 | 1.03E+00 | 9.20E-01 |
| 14 | 8 | 5.23E-01 | 7.20E+00 |
| 15 | 10 | 1.68E-01 | 1.02E+00 |
| 15 | 12 | 1.82E+00 | 1.62E-01 |
| 16 | 14 | 2.92E-01 | 2.55E-01 |
| 16 | 16 | 1.33E+00 | 2.34E+00 |
| 16 | 18 | 4.72E+00 | 5.47E+00 |
| 16 | 20 | 5.43E-01 | 5.08E-01 |
| 17 | 22 | 2.69E-01 | 2.56E-01 |
| 17 | 24 | 3.26E+00 | 2.73E+00 |
| 17 | 26 | 4.57E+01 | 4.23E+00 |
| 17 | 28 | 3.85E+00 | 3.97E+01 |
| 17 | 30 | 6.59E+00 | 3.78E+00 |
| 18 | 32 | 1.63E+00 | 1.15E+00 |
| 18 | 34 | 2.16E+00 | 1.50E+00 |

**Table 4a**. DD response matrix: Doubling to convergence at $N = 34$

| | | $R_f$ | | | $T_n$ | | |
|---|---|---|---|---|---|---|---|
| $n$ | $N$ | $\varepsilon_r(ORI)/\varepsilon_{min}$ | $\varepsilon_r/(W-e)/\varepsilon_{min}$ | $\varepsilon_r(R)/\varepsilon_{min}$ | $\varepsilon_r(ORI)/\varepsilon_{min}$ | $\varepsilon_r(W-e)/\varepsilon_{min}$ | $\varepsilon_r(R)/\varepsilon_{min}$ |
| 2 | 34 | 1.00E+00 | 1.00E+00 | 1.00E+00 | 1.00E+00 | 1.00E+00 | 1.00E+00 |
| 3 | 34 | 1.00E+00 | 1.00E+00 | 1.00E+00 | 1.00E+00 | 1.00E+00 | 1.00E+00 |
| 4 | 34 | 1.29E+00 | 1.00E+00 | 1.93E+00 | 1.29E+00 | 1.00E+00 | 1.93E+00 |
| 5 | 34 | 5.41E+00 | 1.22E+01 | 1.00E+00 | 5.41E+00 | 1.22E+01 | 1.00E+00 |
| 6 | 34 | 2.73E+00 | 1.01E+00 | 1.00E+00 | 2.73E+00 | 1.01E+00 | 1.00E+00 |
| 7 | 34 | 1.59E+00 | 1.56E+01 | 1.00E+00 | 1.59E+00 | 1.56E+01 | 1.00E+00 |
| 8 | 34 | 7.54E+00 | 3.56E+02 | 1.00E+00 | 7.54E+00 | 3.56E+02 | 1.00E+00 |
| 9 | 34 | 1.53E+03 | 1.48E+03 | 1.00E+00 | 1.54E+03 | 1.49E+03 | 1.00E+00 |

**Table 4b**. DDM1 response matrix: Doubling to convergence at $N = 34$

| $n$ | $N$ | $\varepsilon_r(ORI)/\varepsilon_{min}$ | $\varepsilon_r/(W-e)/\varepsilon_{min}$ | $\varepsilon_r(R)/\varepsilon_{min}$ | $\varepsilon_r(ORI)/\varepsilon_{min}$ | $\varepsilon_r(W-e)/\varepsilon_{min}$ | $\varepsilon_r(R)/\varepsilon_{min}$ |
|---|---|---|---|---|---|---|---|
| 2 | 34 | 1.00E+00 | 1.00E+00 | 1.00E+00 | 1.00E+00 | 1.00E+00 | 1.00E+00 |
| 3 | 34 | 1.00E+00 | 1.00E+00 | 1.00E+00 | 1.00E+00 | 1.00E+00 | 1.00E+00 |
| 4 | 34 | 1.00E+00 | 1.13E+00 | 1.23E+00 | 1.00E+00 | 1.13E+00 | 1.23E+00 |
| 5 | 34 | 1.00E+00 | 1.13E+00 | 2.20E+00 | 1.00E+00 | 1.13E+00 | 2.20E+00 |
| 6 | 34 | 1.00E+00 | 2.62E+00 | 9.72E+00 | 1.00E+00 | 2.62E+00 | 9.72E+00 |
| 7 | 34 | 1.00E+00 | 2.98E+02 | 1.69E+03 | 1.00E+00 | 2.99E+02 | 1.70E+03 |

**Table 4c**. DDM2 response matrix: Doubling to convergence at $N = 34$

| $n$ | $N$ | $\varepsilon_r(ORI)/\varepsilon_{min}$ | $\varepsilon_r/(W-e)/\varepsilon_{min}$ | $\varepsilon_r(R)/\varepsilon_{min}$ | $\varepsilon_r(ORI)/\varepsilon_{min}$ | $\varepsilon_r(W-e)/\varepsilon_{min}$ | $\varepsilon_r(R)/\varepsilon_{min}$ |
|---|---|---|---|---|---|---|---|
| 2 | 34 | 1.00E+00 | 1.00E+00 | 1.00E+00 | 1.00E+00 | 1.00E+00 | 1.00E+00 |
| 3 | 34 | 1.00E+00 | 1.00E+00 | 1.00E+00 | 1.00E+00 | 1.00E+00 | 1.00E+00 |
| 4 | 34 | 1.63E+00 | 1.00E+00 | 2.36E+00 | 1.64E+00 | 1.00E+00 | 2.40E+00 |
| 5 | 34 | 1.00E+00 | 5.24E+00 | 1.03E+00 | 1.00E+00 | 4.57E+00 | 1.04E+00 |
| 6 | 34 | 1.00E+00 | 1.80E+00 | 1.65E+00 | 1.00E+00 | 1.87E+00 | 1.60E+00 |
| 7 | 34 | 1.00E+00 | 1.33E+04 | 1.43E+05 | 1.00E+00 | 1.37E+04 | 1.44E+05 |
| 8 | 34 | 1.28E+00 | 1.00E+00 | 2.33E+00 | 1.22E+00 | 1.00E+00 | 2.23E+00 |
| 9 | 34 | 1.00E+00 | 3.35E+06 | 1.99E+05 | 1.00E+00 | 3.35E+06 | 1.99E+05 |
| 10 | 34 | 1.00E+00 | 1.75E+05 | 5.25E+04 | 1.00E+00 | 1.75E+05 | 5.24E+04 |
| 11 | 34 | 1.00E+00 | 4.12E+04 | 3.33E+03 | 1.00E+00 | 4.13E+04 | 3.33E+03 |

## 2. Choice of best response matrix approximation

The convergence profiles just shown provide an efficient way to compare the four response matrix approximations. Tables 4a-d present the doubling convergence and last three entries of the quadrature convergence profiles for the DD, DDM1 and DDM2 approximations. Of particular note is that all response matrix approximations converge at the same quadrature order, however the number of doublings and CPU times differ according to Table 5a. While not shown, all $R_f + T_n$ add to unity to more than 6–places. Moreover, the higher order response matrices approximations (DDM1, DDM2) converge by the original sequence; however, all angular convergence is by *W-e* as well as on the run-up to angular convergence.

This suggest a modified spatial convergence scheme. Rather than begin the spatial convergence at $n = 2$, we begin at $n = 7$ and save the converged doubling order $n_c(N)$ on convergence at the current $N$. We then begin the doubling sequence of the (next) incremented quadrature with the converged doubling of the last converged value minus one, $n_c(N)-1$. This avoids the low order spatial discretizations that we know are inaccurate.

Table 5b shows the improved convergence characteristics for each response matrix approximation for the modified scheme.

**Table 4d**. DD/DDM1/DDM2 response matrix: Quadrature order convergence

| n | N | $\varepsilon_r(ORI)/\varepsilon_r(W\text{-}e)$ | $\varepsilon(ORI)/\varepsilon(W\text{-}e)$ |
|---|---|---|---|
| … | … | … | … |
| 7 | 30 | 7.13E+00 | 3.13E+00 |
| 7 | 32 | 1.69E+00 | 1.01E+00 |
| 7 | 34 | 2.65E+00 | 2.65E+00 |
| … | … | … | … |
| 9 | 30 | 7.21E+00 | 3.19E+00 |
| 9 | 32 | 1.69E+00 | 1.02E+00 |
| 9 | 34 | 2.64E+00 | 2.64E+00 |
| … | … | … | … |
| 7 | 30 | 7.13E+00 | 3.13E+00 |
| 7 | 32 | 1.69E+00 | 1.01E+00 |
| 7 | 34 | 2.65E+00 | 2.65E+00 |
| … | … | … | … |
| 11 | 30 | 7.42E+00 | 3.31E+00 |
| 11 | 32 | 1.72E+00 | 1.05E+00 |
| 11 | 34 | 2.64E+00 | 2.63E+00 |

**Table 5a**. Comparison of Response matrix approximation

| RM | $n_c$ | $N_c$ | Spatial Convergence Mode | Angular Convergence Mode | CPU Time(s) |
|---|---|---|---|---|---|
| BE | 18 | 34 | R | W-e | 7.5 |
| DD | 9 | 34 | R | W-e | 2.1 |
| DD1 | 7 | 34 | ORI | W-e | 1.6 |
| DD2 | 11 | 34 | ORI | W-e | 4.0 |

**Table 5b**. Modified spatial convergence scheme

| RM | $n_c$ | $N_c$ | Spatial Convergence Mode | Angular Convergence Mode | CPU Time(s) |
|---|---|---|---|---|---|
| BE | 31* | 40 | *W-e* | *W-e* | 9.0 |
| DD | 13 | 34 | *ORI* | *W-e* | 1.1 |
| DD1 | 8 | 34 | *ORI* | *W-e* | 0.86 |
| DD2 | 13 | 34 | *ORI* | *W-e* | 1.33 |

* Did not converge

Therefore based on this example for the modified spatial convergence scheme, which seems representative of the numerical intensity evaluations for the 3D pipe, we observe a reduction in CPU time (aside from BE) with the best performance by DDM1. Therefore, we choose the DDM1 representation as the working response matrix approximation going forward.

### 3. Limited precision

To end the section, we push the algorithm to see its limitations. Table 6 shows converged intensities for decreasing desired relative error $\varepsilon_r$. As observed, it seems that precision to only one unit in the $9^{th}$ - place is obtainable before roundoff sets in- which is about the best one can expect for double precision arithmetic. This result points out loss of precision in the response matrix, one of the difficulties of doubling as $h$ becomes small. For this reason, to be on the safe side, we will quote only 7-places of accuracy in the benchmarks to follow, which is realistically all that is required for verification.

**Table 6**. $R_f$ and $T_n$ for decreasing requested relative error

| $\varepsilon_r$ | $R_f$ | $T_n$ | $n_c$ | $N_c$ |
|---|---|---|---|---|
| $10^{-7}$ | **3.275882**609E-01 | **6.724117**386E-01 | 8 | 24 |
| $10^{-8}$ | **3.275825**61E-01 | **6.724117**423E-01 | 8 | 30 |
| $10^{-9}$ | **3.27588253**6E-01 | **6.72411746**4E-01 | 8 | 34 |
| $10^{-10}$ | 3.**275882532**E-01 | **6.72411746**8E-01 | 8 | 50 |
| $10^{-11}$ | **3.275882533**E-01 | **6.72411746**7E-01 | 8 | 46 |
| $10^{-12}$ | **3.275882532**E-01 | **6.72411746**8E-01 | 10 | 64 |

### IV. Benchmarks by doubling

In the demonstration to follow, all tables of reflectance and transmittance of Ref. 11 are reproduced with at least two additional digits of precision. We consider both isotropic and beam sources for semi-infinite and finite length pipes with the DDM1 response matrix approximation. The benchmarks are presented to guide algorithm development and to highlight the doubling method in comparison to all other methods applied to pipe particle transport.

## 1. Half-space duct

The half-space duct for an incident isotopic source is treated as a medium of large width, $Z = 150000$. We monitor convergence only for reflectance in this case as $R_f$ is invariant for large widths. In addition, for each quadrature ($N$) increment, $\Delta Z = 200$ is added to the width so that the sequence element includes the width tending toward infinity as it converges.

### 1.a. Isotropic incidence

Table 7a displays 7-place reflectances whose values are in complete agreement with those of Ref. 11, (Table 2) on rounding to the fifth place.

**Table 7a (Table 2[10])**. Half-space: Isotropic incidence

| $\omega$ | $R_f$ | $n_c$ | $N_c$ |
|---|---|---|---|
| 0.1 | 2.4954290E-02 | 12 | 30 |
| 0.3 | 8.4362038E-02 | 12 | 28 |
| 0.6 | 2.1421124E-01 | 11 | 26 |
| 0.9 | 4.9955168E-01 | 11 | 22 |
| 0.95 | 6.1392428E-01 | 11 | 22 |
| 0.99 | 8.0206397E-01 | 10 | 22 |

### 1.b. Beam incidence

The next half-space case is for beam incidence of orientation $\xi_0$ at the entrance to the pipe and presents a special challenge for the RMCAD method.

The incident intensity, for this case, becomes

$$Y(0,\xi) = \begin{bmatrix} 1 & 0 & 0 \end{bmatrix}^T \delta(\xi - \xi_0). \tag{28a}$$

Usually, the beam source, transformed to the uncollided component, becomes an equivalent volume source added to the transport equation. While this is fine for essentially all numerical methods, it is a disaster for doubling. This is so because a variable source must be part of the response of the fundamental slab of Fig. 1. To obtain an expression for the reflectance and transmittance would therefore require the solution of the transport equation over the fundamental slab which itself would mean application of responses at small thicknesses in order to capture the source variation. More importantly, since the responses are now different, the advantage of doubling is lost.

To avoid this issue one applies a similar, but simpler, procedure than that of the nascent delta function found in Ref. 11, called an inline delta function. In particular, one adds one additional ordinate to the quadrature abscissae list at $\xi_0$ with a weight of $1/\varepsilon$. The intensity entering becomes

$$Y(0,\xi) = \begin{bmatrix} 0 & ... & 0 & 1/\varepsilon & 0 & ... & 0 \end{bmatrix}^T, \tag{28b}$$

where the source strength at $\xi_0$ is set to $1/\varepsilon$. This conserves integration over the incident source distribution. One needs to do nothing more than choose $\varepsilon$. To show the insensitivity to the choice of $\varepsilon$, we consider the cases of $\xi_0 = 0.100503...$ and 7.017924... (corresponding to $\mu_0 = 0.1$ and 0.99 of Ref. 11) for $\omega = 0.1$ and 0.99. As one observes from Table 7b, $\varepsilon$ is any small number less than $10^{-10}$ without further adjustment and without consequence.

**Table 7b**. Variation of $\varepsilon$

| $\varepsilon \backslash (\xi_0, \omega)$ | (0.100503, 0.1) | (0.100503, 0.99) | (7.017924, 0.1) | (7.017924, 0.99) |
|---|---|---|---|---|
| $10^{-5}$ | 4.6340632E-02 | 8.9115641E-01 | 4.6952693E-03 | 5.3809641E-01 |
| $10^{-10}$ | 4.6340072E-02 | 8.9108975E-01 | 4.6952690E-03 | 5.3809632E-01 |
| $10^{-15}$ | 4.6340072E-02 | 8.9108975E-01 | 4.6952690E-03 | 5.3809632E-01 |
| $10^{-20}$ | 4.6340072E-02 | 8.9108975E-01 | 4.6952690E-03 | 5.3809632E-01 |
| $10^{-125}$ | 4.6340072E-02 | 8.9108975E-01 | 4.6952690E-03 | 5.3809632E-01 |

Table 7c shows the corresponding results of Ref. 11 (Table 4), which, again, are in complete agreement on rounding.

**Table 7c (Table 4[10])**. Half-space: Beam incidence

| $\xi_0 \backslash \omega$ | 0.1 | 0.3 | 0.6 | 0.9 | 0.95 | 0.99 |
|---|---|---|---|---|---|---|
| 0.100503 | 4.6340072E-02 | 4.3154330E-02 | 3.6458125E-02 | 2.9258990E-02 | 2.0663694E-02 | 4.6952692E-03 |
| 0.204124 | 1.4939267E-01 | 1.4014448E-01 | 1.2048453E-01 | 9.8792364E-02 | 7.1766494E-02 | 1.6995066E-02 |
| 0.436435 | 3.4391717E-01 | 3.2721137E-01 | 2.9090875E-01 | 2.4881433E-01 | 1.9180203E-01 | 5.0899524E-02 |
| 0.749999 | 6.6203478E-01 | 6.4507273E-01 | 6.0680379E-01 | 5.5860539E-01 | 4.8318257E-01 | 1.8572340E-01 |
| 1.333333 | 7.5903205E-01 | 7.4509553E-01 | 7.1328268E-01 | 6.7218829E-01 | 6.0489097E-01 | 2.8220685E-01 |
| 7.017924 | 8.9108991E-01 | 8.8366801E-01 | 8.6645481E-01 | 8.4344777E-01 | 8.0335527E-01 | 5.3809632E-01 |

## 2. Finite length ducts

The final demonstrations are for finite length pipes with isotropic and beam incidences.

Table 8 gives the reflectance and transmittance for an isotropically incident source for the selected $Z/\rho$ and $\omega$ of Ref. 11, (Tables 5 and 6) for three requested errors $\varepsilon_r$. Since we know that for high precision, the quadrature order will generally be larger than 20; and that for small $\xi_0$, the order will be significantly higher, we can improve upon the efficiency of the algorithm by starting at higher quadratures than 2. Therefore, we begin with a quadrature order of 24 and 64 for $\xi_0 > 0.1$ and $< 0.1$ respectively for Table 8 and 12 and 48 for Table 9.

Except for the first reflectance in Table 8, all results for $\varepsilon_r = 10^{-9}$ are precise to 7 places and are in agreement to the four places of Garcia [11] when rounded. To see that convergence acceleration has an advantage, acceleration was suppressed for $\varepsilon_r = 10^{-9}$. The CPU time to completion was nearly double (83.3*s*) that with acceleration (45.5*s*). In conclusion, acceleration is cost effective in achieving high precision.

Table 9 gives intensity from a beam, which is also in agreement with those of Ref. 11.

**Table 8 (Tables $5^{10}$, $6^{10}$).** Error comparison for finite pipes:

| $\varepsilon_r$ | | $10^{-8}$ | | $10^{-9}$ | | $10^{-10}$ | |
|---|---|---|---|---|---|---|---|
| $\omega$ | $Z/\rho$ | $R_f$ | $T_n$ | $R_f$ | $T_n$ | $R_f$ | $T_n$ |
| 0.1 | 1.0E-01 | 4.31634328E-03 | 9.11789866E-01 | 4.31634332E-03 | 9.11789866E-01 | 4.31634315E-03 | 9.11789866E-01 |
|  | 1.0E+00 | 2.16241639E-02 | 3.98937654E-01 | 2.16241639E-02 | 3.98937654E-01 | 2.16241638E-02 | 3.98937654E-01 |
|  | 1.0E+01 | 2.49538350E-02 | 1.01141298E-02 | 2.49538350E-02 | 1.01141298E-02 | 2.49538350E-02 | 1.01141298E-02 |
| 0.3 | 1.0E-01 | 1.31469907E-02 | 9.20603711E-01 | 1.31469904E-02 | 9.20603712E-01 | 1.31469903E-02 | 9.20603712E-01 |
|  | 1.0E+00 | 7.00779039E-02 | 4.40541235E-01 | 7.00779033E-02 | 4.40541235E-01 | 7.00779027E-02 | 4.40541235E-01 |
|  | 1.0E+01 | 8.43598302E-02 | 1.10500266E-02 | 8.43598302E-02 | 1.10500266E-02 | 8.43598301E-02 | 1.10500266E-02 |
| 9,6 | 1.0E-01 | 2.69112343E-02 | 9.34342413E-01 | 2.69112338E-02 | 9.34342413E-01 | 2.69112338E-02 | 9.34342413E-01 |
|  | 1.0E+00 | 1.59585668E-01 | 5.19389405E-01 | 1.59585666E-01 | 5.19389406E-01 | 1.59585666E-01 | 5.19389406E-01 |
|  | 1.0E+01 | 2.14194934E-01 | 1.48408650E-02 | 2.14194934E-01 | 1.48408650E-02 | 2.14194934E-01 | 1.48408650E-02 |
| 0.9 | 1.0E-01 | 4.13375138E-02 | 9.48742735E-01 | 4.13375133E-02 | 9.48742735E-01 | 4.13375131E-02 | 9.48742735E-01 |
|  | 1.0E+00 | 2.78570551E-01 | 6.27225121E-01 | 2.78570550E-01 | 6.27225121E-01 | 2.78570549E-01 | 6.27225122E-01 |
|  | 1.0E+01 | 4.98283262E-01 | 5.07671522E-02 | 4.98283262E-01 | 5.07671521E-02 | 4.98283262E-01 | 5.07671522E-02 |
| 0.95 | 1.0E-01 | 4.38096450E-02 | 9.51210498E-01 | 4.38096444E-02 | 9.51210499E-01 | 4.38096442E-02 | 9.51210499E-01 |
|  | 1.0E+00 | 3.02372923E-01 | 6.49119315E-01 | 3.02372923E-01 | 6.49119315E-01 | 3.02372921E-01 | 6.49119316E-01 |
|  | 1.0E+01 | 6.07739941E-01 | 8.61168574E-02 | 6.07739941E-01 | 8.61168572E-02 | 6.07739941E-01 | 8.61168574E-02 |
| 1.0 | 1.0E-01 | 4.63017629E-02 | 9.53698237E-01 | 4.63017624E-02 | 9.53698238E-01 | 4.63017621E-02 | 9.53698238E-01 |
|  | 1.0E+00 | 3.27588255E-01 | 6.72411745E-01 | 3.27588255E-01 | 6.72411745E-01 | 3.27588253E-01 | 6.72411747E-01 |
|  | 1.0E+01 | 8.09035563E-01 | 1.90964437E-01 | 8.09035563E-01 | 1.90964437E-01 | 8.09035563E-01 | 1.90964437E-01 |
| CPU Time(s) | | 19.6 | | 45.5 | | 171 | |

**Table 9 (Tables $7^{10}$, $8^{10}$).** Finite pipes: Beam incidence

| $\omega$ | | 0.6 | | 1.0 | |
|---|---|---|---|---|---|
| $\xi_0$ | $Z/\rho$ | $R_f$ | $T_n$ | $R_f$ | $T_n$ |
| 0.100503 | 1.0000E-01 | 1.8117739E-01 | 5.5762177E-01 | 3.1174912E-01 | 6.8825088E-01 |
|  | 1.0000E+00 | 3.2522106E-01 | 1.6219110E-01 | 6.4009549E-01 | 3.5990451E-01 |

|          | 1.0000E+01 | 3.4391471E-01 | 1.8184872E-03 | 9.0660003E-01 | 9.3399970E-02 |
|----------|------------|---------------|---------------|---------------|---------------|
|          | 1.0000E-01 | 8.7797351E-02 | 7.8561822E-01 | 1.5105905E-01 | 8.4894095E-01 |
| 0.204124 | 1.0000E+00 | 3.0662546E-01 | 1.7756880E-01 | 6.1112865E-01 | 3.8887135E-01 |
|          | 1.0000E+01 | 3.2720877E-01 | 1.9221768E-03 | 8.9945787E-01 | 1.0054213E-01 |
|          | 1.0000E-01 | 4.0388388E-02 | 9.0143217E-01 | 6.9489778E-02 | 9.3051022E-01 |
| 0.436435 | 1.0000E+00 | 2.6531103E-01 | 2.1680649E-01 | 5.4254873E-01 | 4.5745127E-01 |
|          | 1.0000E+01 | 2.9090579E-01 | 2.1706787E-03 | 8.8267542E-01 | 1.1732458E-01 |
|          | 1.0000E-01 | 2.3698724E-02 | 9.4216614E-01 | 4.0774983E-02 | 9.5922502E-01 |
| 0.749999 | 1.0000E+00 | 1.9416468E-01 | 3.9364166E-01 | 4.0242802E-01 | 5.9757198E-01 |
|          | 1.0000E+01 | 2.4881083E-01 | 2.5343391E-03 | 8.5960792E-01 | 1.4039208E-01 |
|          | 1.0000E-01 | 1.3499906E-02 | 9.6705270E-01 | 2.3227544E-02 | 9.7677246E-01 |
| 1.333333 | 1.0000E+00 | 1.1290030E-01 | 6.4224806E-01 | 2.3474094E-01 | 7.6525906E-01 |
|          | 1.0000E+01 | 1.9179734E-01 | 3.3268328E-03 | 8.1727090E-01 | 1.8272910E-01 |
|          | 1.0000E-01 | 2.6195567E-03 | 9.9360559E-01 | 4.5071776E-03 | 9.9549282E-01 |
| 7.017924 | 1.0000E+00 | 2.1032720E-02 | 9.3387145E-01 | 4.3644289E-02 | 9.5635571E-01 |
|          | 1.0000E+01 | 5.0601604E-02 | 2.0846355E-01 | 4.3418087E-01 | 5.6581913E-01 |

## DISCUSSION

The revival of an old transport method has been presented and applied to an old transport problem. One observes that the method of adding and doubling wrapped in convergence acceleration is an effective way to solve the 1D transport equation for neutral particle transport in a duct. The following two sections highlight the major differences between the method of doubling and the NDO and ADO methods.

### a. Comparison to the NDO method

The advantage of doubling over NDO primarily rests in numerical performance. Both doubling and NDO methods are about as numerically unsophisticated as a method can be, unlike the ADO method. They are both fully discretized, but differ in the numerical algorithm for solution of the discretized forms. Thus, they both accumulate spatial truncation and angular discretization errors as well as roundoff error. At a fixed quadrature order, the spatial discretization error of, say for DD, order $h^2$ in the intensity, comes from the response matrix approximation of that error order. For the NDO method of the same discretization, the order of error is identical to that of doubling. Hence, as the two algorithms march across the slab, either by doubling or sweeping of the diamond difference equations, the error accumulates. As shown above, $n$ doublings are required to cover the slab by doubling to give the exiting distributions; while, for the NDO method, $2^n$ intervals are required in two sweeps to cover the entire slab over all orientations. Thus, the accumulation of spatial truncation error is significantly less for doubling. This is the source of the efficient computation of doubling and its advantage.

The second advantage of doubling over the NDO method is that it is explicit in the spatial and angular variable; while, NDO requires iteration in the form of sweeps in the directional variable. Essentially, both methods are inverting a large matrix of order $2^n \times 6N$, with the NDO method performing iteratively and the RMCAD method performing efficient explicit matrix inversions of several sub-matrices of order $3N \times 3N$.

Finally, since we use convergence acceleration, there is no need for experimentation to determine the best discretization and quadrature order. Convergence acceleration is essentially a consistent sensitivity study designed to limit to a converged solution. NDO, on the other hand, is a single realization of the numerical solution, which is usually a part of an inconclusive ad hoc sensitivity study.

Table 10 provides a direct comparison between RMCAD and NDO limited to 5-places for comparison to Tables I and II of Ref. 6. First to note is that thicker slabs do not require an excessive numerical effort compared to thinner slabs as seems true for NDO. Moreover, the maximum quadrature for doubling over all cases is 32 compared to 640 for NDO. Doubling required 256 intervals or 8 doublings for all cases, which is greater than that used by NDO, but has minimal impact because of the spatial efficiency of the doubling algorithm. Finally, of note is the greatly reduced overall computational times relative to NDO even adjusting for the slower computer used in Ref. 6.

**b. Comparison to the ADO method**
It is clear that RMCAD produces results as precise as ADO does. There is also no doubt that ADO can generate the 7-place results obtained by doubling presented in the tables of this work. Realistically, the number of correct digits after five becomes academic and usually finds use only in the further scholastic development of benchmarks and consequently may not be the most appropriate measure of the overall value of a numerical transport method.

There are many methods developers of solutions to the transport equation each attempting to outdo the other. In the view of the author, the ADO method has held the lead position as the premier transport solver of the current era, having been successfully applied to nearly every 1D or 1D-like transport equation known to humans by Siewert, Barichello, Garcia and co-workers. While the revival of the doubling method developed in this work is in its early stages, its accuracy through simplicity should be recognized as a challenge to the supremacy of the ADO method.

Central to the ADO method is the determination of eigenvalues and eigenvectors, which has become rather sophisticated over the many years of development of ADO methods. The success of ADO comes in large part form the considerable effort in perfecting their determination. Eigenvalues and eigenvectrors are not required by RMCAD, where the only mathematical operations are matrix arithmetic and inversion also required by ADO. Another drawback of the ADO method is its inconsistency in treating the conservative or near conservative case. This is consequence of the solutions to the homogeneous ADO equations becoming dependence when the eigenvalue approaches zero. In contrast to RMCAD, there is no need to distinguish between cases.

A second measure is efficiency in precision and computational effort. While it is difficult to make meaningful timing comparisons with ADO, Table 11 is included to estimate levels of computational effort for the results compared in the tables of the previous section possibly to aid future comparisons.

**Table 10**. Time comparison with 4-places of Ref. 6 Tables I and II.

| $\omega$ | $Z/\rho$ | $R_f$ | $T_n$ | $N_c$ | *CPU Time*(s) |
|---|---|---|---|---|---|
| 0.1 | 1.0000E-01 | 4.3163E-03 | 9.1179E-01 | 30 | 5.31E-01 |
|  | 1.0000E+00 | 2.1624E-02 | 3.9894E-01 | 18 | 9.38E-02 |
|  | 1.0000E+01 | 2.4954E-02 | 1.0114E-02 | 22 | 1.72E-01 |
| 0.2 | 1.0000E-01 | 8.6982E-03 | 9.1616E-01 | 30 | 5.31E-01 |
|  | 1.0000E+00 | 4.4913E-02 | 4.1883E-01 | 16 | 4.69E-02 |
|  | 1.0000E+01 | 5.2832E-02 | 1.0514E-02 | 24 | 2.19E-01 |
| 0.3 | 1.0000E-01 | 8.6982E-03 | 9.1616E-01 | 30 | 5.31E-01 |
|  | 1.0000E+00 | 4.4913E-02 | 4.1883E-01 | 16 | 4.69E-02 |
|  | 1.0000E+01 | 5.2832E-02 | 1.0514E-02 | 24 | 2.19E-01 |
| 0.4 | 1.0000E-01 | 8.6982E-03 | 9.1616E-01 | 30 | 5.16E-01 |
|  | 1.0000E+00 | 4.4913E-02 | 4.1883E-01 | 16 | 4.69E-02 |
|  | 1.0000E+01 | 5.2832E-02 | 1.0514E-02 | 24 | 2.19E-01 |
| 0.5 | 1.0000E-01 | 2.2252E-02 | 9.2969E-01 | 30 | 5.31E-01 |
|  | 1.0000E+00 | 1.2709E-01 | 4.9050E-01 | 16 | 6.25E-02 |
|  | 1.0000E+01 | 1.6302E-01 | 1.2949E-02 | 20 | 1.09E-01 |
| 0.6 | 1.0000E-01 | 2.6911E-02 | 9.3434E-01 | 30 | 5.16E-01 |
|  | 1.0000E+00 | 1.5959E-01 | 5.1939E-01 | 14 | 4.69E-02 |
|  | 1.0000E+01 | 2.1419E-01 | 1.4841E-02 | 20 | 1.25E-01 |
| 0.7 | 1.0000E-01 | 3.1644E-02 | 9.3907E-01 | 28 | 3.91E-01 |
|  | 1.0000E+00 | 1.9530E-01 | 5.5144E-01 | 14 | 3.12E-02 |
|  | 1.0000E+01 | 2.7850E-01 | 1.8399E-02 | 26 | 3.12E-01 |
| 0.8 | 1.0000E-01 | 3.6452E-02 | 9.4387E-01 | 28 | 3.91E-01 |
|  | 1.0000E+00 | 2.3474E-01 | 5.8717E-01 | 14 | 4.69E-02 |
|  | 1.0000E+01 | 3.6502E-01 | 2.6434E-02 | 22 | 1.72E-01 |
| 0.9 | 1.0000E-01 | 4.1338E-02 | 9.4874E-01 | 28 | 3.75E-01 |
|  | 1.0000E+00 | 2.7857E-01 | 6.2722E-01 | 16 | 4.69E-02 |
|  | 1.0000E+01 | 4.9828E-01 | 5.0767E-02 | 32 | 6.72E-01 |
|  | 1.0000E-01 | 4.5802E-02 | 9.5320E-01 | 22 | 1.72E-01 |
| 0.99 | 1.0000E+00 | 3.2243E-01 | 6.6764E-01 | 16 | 4.69E-02 |
| | 1.0000E+01 | 7.5402E-01 | 1.5759E-01 | 16 | 6.25E-02 |
| | 1.0000E-01 | 4.6302E-02 | 9.5370E-01 | 22 | 1.56E-01 |
| 1.0 | 1.0000E+00 | 3.2759E-01 | 6.7241E-01 | 16 | 4.69E-02 |
| | 1.0000E+01 | 8.0904E-01 | 1.9096E-01 | 16 | 4.69E-02 |

**Table 11.** Timing Comparison

| $\varepsilon_r$ | $10^{-9}$ | | | $10^{-5}$ | | |
|---|---|---|---|---|---|---|
| Table | Time(s) | $n_c$ | $N_c$ | Time(s) | $n_c$ | $N_c$ |
| Table 7a ($2^{10}$) | 2.6 | 12 | 36 | 0.34 | 11 | 18 |
| Table 7c ($4^{10}$) | 19.2 | 16 | 34 | 3.1 | 16 | 20 |
| Table 8 ($5^{10}$ and $6^{10}$) | 45.5 | 9 | 68 | 12.9 | 8 | 34 |
| Table 9 ($7^{10}$ and $8^{10}$) | 108 | 10 | 60 | 28.9 | 8 | 46 |
| Table 10 ($I^6$ and $II^6$) | -- | -- | -- | 7.5 | 8 | 30 |

A second comparison comes from Table 2 of Ref. 11, which is the run-up to convergence in quadrature order. Table 12 shows the corresponding doubling run-up. In comparison, convergence is generally at (shaded) or within one quadrature increment (boxed-shaded entry) of ADO indicating similarity of the performance of the two methods. Hopefully, future head to head comparisons will be made to quantify their performance.

**Table 12**. Convergence in $N$: Semi-infinite medium with isotropic incidence

| $N\backslash\omega$ | 0.1 | 0.3 | 0.6 | 0.9 | 0.95 | 0.99 |
|---|---|---|---|---|---|---|
| 2 | 1.77923E-02 | 5.72659E-02 | 1.34582E-01 | 2.83622E-01 | 3.35625E-01 | 3.98948E-01 |
| 3 | 2.94203E-02 | 9.87594E-02 | 2.47361E-01 | 5.64763E-01 | 6.93705E-01 | 9.63050E-01 |
| 4 | 2.46761E-02 | 8.29732E-02 | 2.09366E-01 | 4.87169E-01 | 5.98343E-01 | 7.77048E-01 |
| 5 | 2.52017E-02 | 8.51309E-02 | 2.15859E-01 | 5.02484E-01 | 6.17358E-01 | 8.06844E-01 |
| 6 | 2.49562E-02 | 8.43344E-02 | 2.14039E-01 | 4.99052E-01 | 6.13327E-01 | 8.01130E-01 |
| 7 | 2.49692E-02 | 8.44052E-02 | 2.14293E-01 | 4.99671E-01 | 6.14063E-01 | 8.02254E-01 |
| 8 | 2.49564E-02 | 8.43658E-02 | 2.14210E-01 | 4.99534E-01 | 6.13903E-01 | 8.02044E-01 |
| 9 | 2.49558E-02 | 8.43658E-02 | 2.14217E-01 | 4.99557E-01 | 6.13929E-01 | 8.02077E-01 |
| 10 | 2.49548E-02 | 8.43633E-02 | 2.14212E-01 | 4.99552E-01 | 6.13923E-01 | 8.02066E-01 |
| 11 | 2.49546E-02 | 8.43628E-02 | 2.14212E-01 | 4.99552E-01 | 6.13924E-01 | 8.02065E-01 |
| 12 | 2.49545E-02 | 8.43624E-02 | 2.14212E-01 | 4.99552E-01 | 6.13924E-01 | 8.02064E-01 |
| 13 | 2.49544E-02 | 8.43623E-02 | 2.14212E-01 | 4.99552E-01 | 6.13924E-01 | 8.02064E-01 |
| 14 | 2.49544E-02 | 8.43622E-02 | 2.14211E-01 | 4.99552E-01 | 6.13924E-01 | 8.02064E-01 |
| 15 | 2.49543E-02 | 8.43621E-02 | 2.14211E-01 | 4.99552E-01 | 6.13924E-01 | 8.02064E-01 |
| 16 | 2.49543E-02 | 8.43621E-02 | 2.14211E-01 | 4.99552E-01 | 6.13924E-01 | 8.02064E-01 |
| 17 | 2.49543E-02 | 8.43621E-02 | 2.14211E-01 | 4.99552E-01 | 6.13924E-01 | 8.02064E-01 |
| 18 | 2.49543E-02 | 8.43621E-02 | 2.14211E-01 | 4.99552E-01 | 6.13924E-01 | 8.02064E-01 |
| 19 | 2.49543E-02 | 8.43621E-02 | 2.14211E-01 | 4.99552E-01 | 6.13924E-01 | 8.02064E-01 |
| 20 | 2.49543E-02 | 8.43621E-02 | 2.14211E-01 | 4.99552E-01 | 6.13924E-01 | 8.02064E-01 |
| 21 | 2.49543E-02 | 8.43620E-02 | 2.14211E-01 | 4.99552E-01 | 6.13924E-01 | 8.02064E-01 |

**c. Limitations of RMCAD**

While the performance of RMCAD is excellent for the cases presented in this work, the method is not completely general. In particular, if there is spatial variation of a volume source, the advantage of doubling is lost as mentioned above. In addition the method becomes unstable for extremely small slab widths, which is only an issues if extreme precision greater than 9–places is desired.

**CONCLUSION**

There should be little doubt that doubling is a viable method of solution for transport in a duct and other transport scenarios. It is more straightforward than ADO and possibly more efficient and should be held in the same regard as ADO as a superior method of solution. The answer to which method, RMCAD or ADO is best however is mostly a matter of taste and will require further inquiry.

**Acknowledgement**: I gratefully acknowledge the help of R.D.M Garcia in the preparation of this manuscript by making his preprint of Ref. 11 available.